\documentclass[a4paper]{jpconf}
\usepackage{graphicx}
\begin{document}
\title{How to (Path-) Integrate by Differentiating\footnote{Based on a presentation given by AK at the 
Seventh International Workshop DICE2014 in Castiglioncello (Italy), September 15-19, 2014. Appeared as A. Kempf, D.M. Jackson, A.H. Morales, J. Phys.: Conf. Ser. 626 012015 (2015)
}}

\author{Achim Kempf$^1$, David M. Jackson$^2$,  Alejandro H. Morales$^3$}

\address{$ $ \\ $^1$Departments of Applied Mathematics and Physics\\$^2$Department of Combinatorics and Optimization\\
University of Waterloo, Ontario N2L 3G1, Canada,\\ 
$^3$Department of Mathematics, UCLA, Los Angeles, CA 90095, USA}

\begin{abstract}
Recently, it was found that a new set of simple techniques allow one to conveniently express ordinary integrals through differentiation. These techniques add to the general toolbox for integration and integral transforms such as the Fourier and Laplace transforms. The new methods also yield new perturbative expansions when the integrals cannot be solved analytically. Here, we add new results, for example, on expressing the Laplace transform and its inverse in terms of derivatives. The new methods can be used to express path integrals in terms of functional differentiation, and they also suggest new perturbative expansions in quantum field theory. 

\end{abstract}

\section{Introduction}
Path integrals over fields, such as that for a scalar field,
\begin{equation}
Z[J] = \int e^{iS[\phi] + i\int d^4x~J(x)\phi(x)}D[\phi]\label{pi},
\end{equation}
which can also be viewed as the Fourier transform of $e^{iS[\phi]}$, are at the very heart of quantum field theory. In spite of the seeming simplicity of such functional integrals, or functional Fourier transforms, they are notoriously difficult to define and evaluate, see, e.g., \cite{weinberg}-\cite{enc}. 

Compared to functional integration, functional differentiation is relatively straightforward and it is of course widely used, for example, in extremization problems such as those arising from variational principles. This has motivated recent work, \cite{jpa}, in which it was shown that in certain circumstances functional integration can be expressed through functional differentiation. This also showed that ordinary integrals can be conveniently expressed in terms of differentiation. This includes proper and improper integrals and also integral transforms such as the Fourier and Laplace transforms. The new methods of \cite{jpa} are not meant to replace existing integration methods; they merely add to the toolbox of techniques for integrals and integral transforms. The new tools have the advantage that when they apply they are often quick and simple to use, work naturally with distributions and can even sidestep the need for contour integrations. Further, when integrals cannot be fully carried out, the new methods suggest new perturbative approaches.   

Here, we will review the new tools, add simpler proofs and we also add some new results. 

\section{Integration by differentiation}
In \cite{jpa}, the following representations of integration through differentiation were shown: 
\begin{eqnarray}
\int_a^b f(x)~dx & = & \lim_{\epsilon\rightarrow 0} f(\partial_\epsilon)~\frac{e^{\epsilon b}-e^{\epsilon a}}{\epsilon}\label{ab}\\
\int_{-\infty}^\infty f(x)~dx & = & \lim_{\epsilon\rightarrow 0}2\pi f(-i\partial_\epsilon)~\delta(\epsilon)\label{oldint}\\
\int_{-\infty}^\infty f(x)~dx & = & 2\pi \delta(i\partial_\epsilon)~f(\epsilon)\label{w}
\end{eqnarray}
(Notice that, as was shown in \cite{jpa}, in Equ.\ref{w} it is not necessary to take $\epsilon$ to a fixed value because the $\epsilon$-dependence drops out.)
Here, $f(\nu\partial_\epsilon)$ with $\nu\in\{\pm 1, \pm i\}$ means in the simplest cases that when $f$ is expanded in a MacLaurin series, its argument is to be replaced by $\nu\partial_\epsilon$. More generally, $f$ may be interpreted as a function on the spectrum of the operator $\partial_\epsilon$. We will address this point again after Equ.\ref{pt}. In particular, when $f(\nu\partial_\epsilon)$ acts on one of its eigenfunctions, $e^{\epsilon a}$, we have: 
\begin{equation}
f(\nu\partial_\epsilon)~ e^{\epsilon a} = f(\nu a) e^{\epsilon a} \label{dmj5}
\end{equation}
While we will give proofs below that apply to fairly large classes of functions $f$, the above equations also hold in a distributional sense. The conditions and exact size of the space of generalized functions for which the above equations hold is not yet known.   

Let us now add the following equations which follow straightforwardly from Equ.\ref{ab} by taking the limits $a\to 0, b\to \infty$ and $a\rightarrow -\infty,b \to 0$ respectively:
\begin{eqnarray}
\int_0^\infty f(x)~dx & = & \lim_{\epsilon\rightarrow 0^+} f(-\partial_\epsilon)~\frac{1}{\epsilon}\label{new1}\\
\int_{-\infty}^0 f(x)~dx & = & \lim_{\epsilon\rightarrow 0^+} f(\partial_\epsilon)~\frac{1}{\epsilon}\\
\int_{-\infty}^\infty f(x)~dx & = & \lim_{\epsilon\rightarrow 0^+} \left(f(\partial_\epsilon)+f(-\partial_\epsilon)\right)~\frac{1}{\epsilon}
\end{eqnarray}

\section{Examples of integration by differentiation}
In \cite{jpa}, the aim was to derive new methods for integrals and integral transform for applications to quantum field theoretic path integrals. 
Here, we will look at these integration methods' general utility. To this end, in order to illustrate the above tools let us perform an integral that is usually considered nontrivial in the sense that it is usually done using contour integration. Namely, using the new Equ.\ref{new1}:
\begin{eqnarray}
\int_0^\infty \frac{\sin(x)}{x} ~dx & = & \lim_{\epsilon\rightarrow 0^+}~\frac{1}{2i}\left(e^{-i\partial_\epsilon}-e^{i\partial_\epsilon}\right)\frac{1}{-\partial_\epsilon}~
\frac{1}{\epsilon}\\
& = & \lim_{\epsilon\rightarrow 0^+}~\frac{-1}{2i}\left(e^{-i\partial_\epsilon}-e^{i\partial_\epsilon}\right)\left( \ln(\epsilon)+c\right)\\
& = & \lim_{\epsilon\rightarrow 0^+}~\frac{-1}{2i}\left(\ln(\epsilon -i) +c -\ln(\epsilon +i)-c\right)\\
& = & \frac{-1}{2i}\left(\frac{-i\pi}{2}-\frac{i\pi}{2}\right) = \frac{\pi}{2}
\end{eqnarray}
Here, we used that the antiderivative of $1/\epsilon$ is the logarithm with an integration constant $c$, and we used that the exponentiated derivative translates: $e^{a\partial_\epsilon}g(\epsilon) = g(\epsilon+a)$. 
For comparison, we now perform a similar integral using Equ.\ref{oldint}:
\begin{eqnarray}
\int_{-\infty}^\infty \frac{\sin(x)}{x}~ dx & = & 2 \pi \lim_{\epsilon\rightarrow 0} \frac{1}{ 2i}\left(e^{\partial_\epsilon} -e^{-\partial_\epsilon}\right)\frac{1}{-i\partial_\epsilon} ~\delta(\epsilon)\\
 & = & \pi \lim_{\epsilon\rightarrow 0} \left(e^{\partial_\epsilon}-e^{-\partial_\epsilon}\right) \left(\Theta(\epsilon) +c'\right)\\
 & = & \pi \lim_{\epsilon\rightarrow 0} \left(\Theta(\epsilon+1) +c' -\Theta(\epsilon-1)-c'\right)\\
 & = & \pi
\end{eqnarray} 
Here, $\Theta$ is the Heaviside function and $c'$ is an integration constant. As tools, we used that $\partial_\epsilon^{-1}$ produces the antiderivative and that $e^{a\partial_\epsilon}$, via Taylor's theorem, is the  operator that translates by $a$. On the Taylor expansion of distributions such as the Dirac Delta and the Heaviside function, see, e.g., \cite{td}. Many integrals can quickly be done analogously, for example: 
\begin{eqnarray}
\int_{-\infty}^\infty \frac{\sin^5(x)}{x} ~dx & = & 3\pi/8\\
\int_{-\infty}^\infty \frac{\sin^2(x)}{x^2}~dx & = & \pi \\ 
\int_{-\infty}^\infty \frac{(1-\cos(tx))}{x^2}~ dx & = & \pi \vert t\vert \\ 
\int_{-\infty}^\infty x^2 ~\cos(x) ~e^{-x^2} ~dx & = & \sqrt{\pi} e^{-1/4}/4 \label{pt}
\end{eqnarray}
Tools that can be useful for more complicated integrals are, e.g., 
\begin{equation}
e^{a\partial_\epsilon^2}~\delta(\epsilon) ~= ~ \frac{1}{\sqrt{4\pi a}}~e^{\frac{-\epsilon^2}{4a}} 
\end{equation}
(see \cite{jpa}) and for $k\ge 0$:
\begin{equation}
\frac{1}{k} ~=~ \int_0^\infty e^{-w k}~dw \label{dummy}
\end{equation}
Let us now consider the integral $\int_{-\infty}^\infty \frac{1}{1+x^2}~dx$. This is not a difficult integral as it could be done via substitution. But it is instructive to consider how the new methods can be applied here since, unlike in the previous examples, this integrand is not an entire function because it possesses poles at $\pm i$. It therefore does 
not possess a power series expansion that converges on all of the real line. Instead of trying to integrate over the real line at once, using Equ.\ref{oldint}, let us therefore split the integral into two:
$\int_{-\infty}^\infty\frac{1}{1+x^2}~dx = \int_{-1}^1\frac{1}{1+x^2}~dx + \int_{I\!\!R \backslash [-1,1]}\frac{1}{1+x^2}~dx$.
A change of variable, $x\to1/x$, in the second term shows that it equals the first term. On the interval $[-1,1]$, the integrand has a convergent Maclaurin series, namely the geometric series. We can now use the new integration method of Equ.\ref{ab}:
\begin{eqnarray}
\int_{-\infty}^\infty\frac{1}{1+x^2}~dx = 2 \int_{-1}^1\frac{1}{1+x^2}~dx
 & = & 2~\lim_{\epsilon\to 0}~\frac{1}{1+\partial_\epsilon^2}~\frac{e^\epsilon-e^{-\epsilon}}{\epsilon}\\
  & = & 2~\lim_{\epsilon\to 0}~\sum_{n=0}^\infty (-\partial_\epsilon^2)^n~\frac{e^\epsilon-e^{-\epsilon}}{\epsilon}\\
   & = & 2~\lim_{\epsilon\to 0}~\sum_{n=0}^\infty (-1)^n \partial_\epsilon^{2n}~
   \sum_{r=0}^\infty\frac{2}{(2r+1)!}~\epsilon^{2r}\\
   & = & 4~\lim_{\epsilon\to 0}~\sum_{n=0}^\infty (-1)^n \partial_\epsilon^{2n}~
   \sum_{r=n}^\infty ~\frac{\epsilon^{2r-2n}}{(2r+1)(2r-2n)!} \\
   & = & 4~\sum_{n=0}^\infty (-1)^n~\frac{1}{2n+1} = 4 ~\arctan(1)= \pi
\end{eqnarray}
As an example of an integral with an ambiguity that needs to be fixed through a pole prescription, let us consider the integral $\int_{-1}^1 \frac{1}{x}~dx$, which can of course take any value, depending on the choice of how the pole is to be approached. This choice also manifests itself in the new methods, namely as a choice of integration constant. Applying Equ.\ref{ab}: 
\begin{eqnarray}
\int_{-1}^1 \frac{1}{x}~dx & = & \lim_{\epsilon\rightarrow 0} \frac{1}{\partial_\epsilon}~\frac{e^{\epsilon b}-e^{\epsilon a}}{\epsilon}\\
 & = & \lim_{\epsilon\to 0} \left(\mbox{Ei}(\epsilon)-\mbox{Ei}(-\epsilon) + \mbox{integration constant}\right)
\end{eqnarray}
Here, $Ei$ is the exponential integral function. We close this section with a quick derivation of Equ.\ref{ab}:
\begin{eqnarray}
\int_a^b f(x)~dx & = & \lim_{\epsilon\rightarrow 0}\int_a^b f(x) e^{-\epsilon x}~dx\\
 & = & \lim_{\epsilon\rightarrow 0}~f(-\partial_\epsilon)\int_a^b e^{-\epsilon x}~dx\\
 & = & \lim_{\epsilon\rightarrow 0}~f(-\partial_\epsilon)~\frac{e^{\epsilon b}-e^{\epsilon a}}{\epsilon}
\end{eqnarray}

\section{Forward and inverse Laplace transforms through differentiation}
As was shown in \cite{jpa}, the representation of integration over the real line given in Equ.\ref{oldint} is a special case of a new representation of Fourier transformation by differentiation. Namely, with the following definition of the Fourier transform $F[f]$ of a function $f$,
\begin{equation} 
F[f](x) = \frac{1}{\sqrt{2\pi}} \int_{-\infty}^\infty e^{ix y}~ f(y)~dy,
\end{equation}
we have the new representation:
\begin{equation} 
F[f](x) = \sqrt{2\pi} ~f(-i\partial_x)~\delta(x) \label{fou}
\end{equation}
It is straightforward to prove, see \cite{jpa}, that Equ.\ref{fou} holds for the basis of plane waves, and they span the function space. This means that the new representation for the integral over the real line given in Equ.\ref{oldint} follows from Equ.\ref{fou} because integration is the zero frequency limit of the Fourier transform.

Our aim now is to show that our new Equ.\ref{new1} is, similarly, the ``zero-frequency" special case of a  new and very simple representation of the Laplace transform, expressed entirely in terms of differentiation. Consider the usual definition of the Laplace transform, $L[f]$, of a function $f$, for $x>0$:
\begin{equation}
L[f](x) = \int_0^\infty f(y) ~e^{-xy}~dy
\end{equation}

\noindent We claim that: 
\begin{eqnarray}
L[f](x) & = & f(-\partial_x)~\frac{1}{x}\label{lap} \\
L^{-1}[f](x) & = & f(\partial_x)~\delta(x)\label{invlap}
\end{eqnarray}
For example, using Equ.\ref{lap}, we can immediately read off the Laplace transforms of monomials, $f(x)=x^n$: 
\begin{equation}
L[f](x) = (-\partial_x)^n~\frac{1}{x} = \frac{n!}{x^{n+1}}
\end{equation}
The new representation, Equ.\ref{invlap}, of the inverse Laplace transform may be particularly useful in practice. This is because, with the usual methods, the inverse Laplace transform is somewhat tedious as it involves analytic continuation.  Equ.\ref{invlap} can be comparatively straightforward to evaluate. For example, let us calculate the inverse Laplace transform of the function $f(x)=1/(x-a)$. Using Equs.\ref{dummy},\ref{invlap}  we obtain:
\begin{eqnarray}
L^{-1}[f](x) & = & \frac{1}{\partial_x-a}~\delta(x)\\
 & = & \int_0^\infty e^{-w(\partial_x-a)}~dw~\delta(x)\\
 & = & \int_0^\infty e^{aw}~\delta(x-w)~dw = e^{ax}
\end{eqnarray}
\section{Proofs for the forward and inverse Laplace transforms}
Equ.\ref{lap} for the forward Laplace transform is easy to derive:
\begin{equation}
L[f](x) = \int_0^\infty e^{-xy}~f(y)~dy = f(-\partial_x)\int_0^\infty e^{-xy}~dy = f(-\partial_x)~\frac{1}{x}
\end{equation}
As a consequence, we have also proven Equ.\ref{new1}, namely as the special case $x\rightarrow 0^+$. 
Let us now show that Equ.\ref{invlap} indeed inverts the action of $L$, i.e., that  $L^{-1} \circ L = id$:
\begin{eqnarray}
L^{-1}\circ L [f] (x) & = & L^{-1}\left[f(-\partial_x) \frac{1}{x}\right] = \lim_{\epsilon\rightarrow 0} L^{-1}\left[f(-\partial_\epsilon)~\frac{1}{\epsilon+x}\right] \\
 & = & \lim_{\epsilon\rightarrow 0} f(-\partial_\epsilon)~L^{-1}\left[\frac{1}{\epsilon+x}\right] 
 = \lim_{\epsilon\rightarrow 0} f(-\partial_\epsilon)~\frac{1}{\epsilon+\partial_x}~\delta(x)  \\
 & = & \lim_{\epsilon\rightarrow 0} f(-\partial_\epsilon)~\int_0^\infty e^{-(\epsilon +\partial_x) w}~dw~\delta(x) \label{dmj32}\\
  & = & \lim_{\epsilon\rightarrow 0} \int_0^\infty f(-\partial_\epsilon)~e^{-\epsilon w} ~\delta(x-w) ~dw\\
  & = & f(x)
\end{eqnarray}
In Equ.\ref{dmj32}, we used Equ.\ref{dummy} and the final step used Equ.\ref{dmj5}.

\section{Example: trace of the heat kernel}
A prominent occurrence of the Laplace transform in quantum field theory is in the context of the trace of the heat kernel, see, e.g., \cite{weinberg,Davies,AK}:
\begin{equation}
h(t) = \sum_n e^{-\lambda_n t}
\end{equation}
Here, the $\lambda_n$ are the (assumed positive) eigenvalues of the heat operator. Given the function $h(t)$ it is possible to recover the spectrum $\{\lambda_n\}$ that generated it. With the new methods, this is particularly simple to see. To this end, let us introduce the spectral comb function:
\begin{equation}
s(\lambda) = \sum_n \delta(\lambda -\lambda_n)
\end{equation}
Clearly, the trace of the heat kernel, $h(t)$, is its Laplace transform: 
\begin{equation}
h(t) =   \sum_n e^{-\lambda_n t} = \int_0^\infty e^{-\lambda t} ~s(\lambda) = L[s](t)
\end{equation}
Using the new representation for the inverse Laplace transform, Equ.\ref{invlap}, we can invert this relation straightforwardly: \begin{eqnarray}
L^{-1}[h](\lambda) & = & L^{-1}\left[\sum_n e^{-\lambda_n t}\right](\lambda) = \sum_n e^{-\lambda_n \partial_\lambda}~\delta(\lambda)
 = \sum_n \delta(\lambda-\lambda_n) = s(\lambda) 
\end{eqnarray}

\section{Outlook}
The methods we have presented here are meant as simple but useful additions to the toolbox for integration and for integral transforms such as the Fourier and Laplace transforms. Being general purpose methods, they may possess applications beyond physics, e.g., to communication engineering and signal processing, \cite{jpa,puetteretal,image}. Within mathematics, there is the possibility that these methods may be extensible covariantly to integration over higher-dimensional curved manifolds. There, a new relationship between integration and differentiation could yield new insights into Stokes' theorem. Here, in one dimension, Stokes' theorem is simply the first fundamental theorem of calculus. We have for the derivative, $f'(x)=df(x)/dx$, as is straightforward to verify: 
\begin{equation}
f'(\partial_\epsilon) = f(\partial_\epsilon)\epsilon - \epsilon f(\partial_\epsilon)
\end{equation}
With this and Equ.\ref{ab}, the first fundamental theorem of calculus takes the form:
\begin{eqnarray}
\int_a^b f'(x) ~dx & = & \lim_{\epsilon\rightarrow 0} \left(f(\partial_\epsilon)\epsilon-\epsilon f(\partial_\epsilon)\right)~\frac{e^{\epsilon b} - e^{\epsilon a}}{\epsilon}\\
 & = & \lim_{\epsilon\rightarrow 0}\left(f(b)~e^{\epsilon b}-f(a)e^{\epsilon a} -\epsilon f(\partial_\epsilon) \frac{e^{\epsilon b} - e^{\epsilon a}}{\epsilon}\right)\\
 & = & f(b)-f(a) -\lim_{\epsilon\rightarrow 0}\epsilon \int_a^b f(x)~dx = f(b)-f(a)
\end{eqnarray}
The challenge is to generalize the present methods to higher dimensions so that, for example, a new perspective into the algebraic workings of boundary and coboundary operators may be gained. For the algebraic background for formal power series, see, for example, \cite{jackson} and for algebraic integration in the context of quantum groups, see \cite{KM}. 

We started this paper by considering the analytical difficulties of quantum field theoretical path integrals. In this context, the new methods could inspire, for example, new approaches to perturbative expansions. Indeed, as was shown in \cite{jpa}, integration by differentiation methods yield not only the small coupling expansion but also naturally the strong coupling expansion. In general, the Dirac Deltas in integration-by-differentiation equations such as Equs.\ref{oldint},\ref{w}, can be regularized in multiple ways, such as the Gaussian and sinc regularisations. Each choice yields another kind of perturbative expansion. 

Also, the new integration methods presented here may provide a new perspective on anomalies in quantum field theories. This is because anomalies in quantum field theory originate in nontrivial transformation properties of the measure in the path integral, see e.g., \cite{fujikawa}, and the new methods may provide a new perspective on this phenomenon. Finally, the question arises whether there exists a generalization that unifies both bosonic and fermionic integration, by expressing both similarly through differentiation.  
\medskip\newline
\noindent \bf Acknowledgement: \rm The authors are grateful for valuable feedback from Michael Trott of Wolfram Alpha LLC. AK and DMJ acknowledge support from NSERC's Discovery program. 
\medskip\newline
\section*{References}

\end{document}